\documentclass[pra, twocolumn, floatfix,superscriptaddress]{revtex4}
\usepackage{graphicx}
\usepackage{epstopdf}
\usepackage{amsmath, amsfonts, amssymb, bm}
\usepackage{hyperref}

\begin{document}
\title{Collective dynamics in a laser-pumped mixture of two atomic ensembles}
\author{Luling Jin}
\email{luling.jin@mpi-hd.mpg.de} 
\affiliation{Max-Planck-Institut f\"ur Kernphysik, Saupfercheckweg 1, D-69117 Heidelberg, Germany}
\affiliation{Institute of Applied Physics, Academy of Sciences of Moldova, Academiei str. 5, MD-2028 Chi\c{s}in\u{a}u, Moldova}

\author{J\"{o}rg \surname{Evers}}
\affiliation{Max-Planck-Institut f\"ur Kernphysik, Saupfercheckweg 1, D-69117 Heidelberg, Germany}

\author{Mihai A. Macovei}
\affiliation{Max-Planck-Institut f\"ur Kernphysik, Saupfercheckweg 1, D-69117 Heidelberg, Germany}
\affiliation{Institute of Applied Physics, Academy of Sciences of Moldova, Academiei str. 5, MD-2028 Chi\c{s}in\u{a}u, Moldova}

\date{\today}

\begin{abstract}
We investigate the quantum dynamics of an atomic mixture composed of two multi-atom ensembles. Each ensemble
is driven separately by a coherent laser field, respectively, and dampens via the interactions with the environmental
vacuum electromagnetic field reservoir. We find that, due to the photon exchange among the two components, long-time
excitation oscillations appear, which may be significantly longer than the inverse lifetime of a single emitter.
Furthermore, few-atom ``jumps'' to the excited state occur as function of the parameter characterizing the inter-components interactions around a certain working point.
\end{abstract}

\maketitle


\section{Introduction}

In collective atomic systems, nearby atoms can interact with each other via virtual or real photons mediated by the vacuum field~\cite{Agarwal,Ficek,FicekPhysRep}.
The dipole-dipole interaction is known to substantially modify the quantum dynamics. An archetype model in this field is the Dicke model~\cite{DickePhysRev.93.99},
which predicts the possibility of superradiance and subradiance, i.e., substantial enhancements or reductions of the total spontaneous emission rate. Not surprisingly, the
spontaneous emission in collective atomic systems has received considerable attention~\cite{PhysRevLett.76.2049,Eberly,andreev,puri,jump,PhysRevA.45.3242,PhysRevA.74.053803,PhysRevA.87.053837}.
Particularly, a strong suppression of the scattered light with respect to the noninteracting atom case was observed very recently \cite{supp}.
The presence of collective effects tends to decrease the sensitivity in the precise measurement of time using atomic transitions. This weakness is proved to
be removable lately by a redesign of the Ramsey-pulse sequence~\cite{ramsey}.
Meanwhile the research on the collective spontaneous emission extends to solid ensembles, such as an ensemble of quantum dots~\cite{PhysRevB.86.245322,PhysRevA.88.043807}, where the excitation can be transferred via charge tunneling~\cite{Bayer19012001},
long-range radiative interaction, or dipole-dipole interaction~\cite{PhysRevB.71.045335,qd_sup}.
Also, an ensemble of Rydberg atoms has strong dipole-dipole interactions and long radiative lifetimes~\cite{Ryd1} and this makes it interesting for the
research of many-body quantum physics~\cite{Ryd2,Ryd3,Ryd4,Ryd5}.
Other types of cooperative effects are observed in cold matter. For instance, in an ultracold boson ensemble trapped in a one-dimensional harmonic
confinement, the quantum breathing dynamics was investigated from few- to many-body systems~\cite{bos}. One interesting aspect of multi-atom ensembles is that they may exhibit collective ``jumps'', i.e., they may sensitively depend on certain control parameters
and change the steady state of the system considerably  already for small variations in this parameter~\cite{andreev,puri,jump}.

Qualitatively new features may arise if the ensemble comprises two or more subensembles. This can be realized, e.g., as a mixture of two atom species
or via spatially separated, yet coupled, ensembles. For example, with two atomic ensembles in a cavity, a cooperatively enhanced index of refraction
without absorption~\cite{mk} or electromagnetically-induced-transparency-like phenomena~\cite{sun} were reported. It is also possible to synchronize
the phase dynamics of two mesoscopic ensembles  through their collective coupling to an optical cavity~\cite{holl} or to create steady-state entanglement
between two distant atomic ensembles at room temperature~\cite{ent}. Recently it was also shown that two far separated subgroups of slow two-level
atoms in a weak resonant laser field can be entangled if they are big enough~\cite{ent2}. Finally, the dipole-dipole interactions among two Rydberg-blockaded
atom clouds break up the superatoms by removing the excitations from the clouds~\cite{block}.

Here, we investigate an atomic mixture composed of two multi-atom ensembles. The two ensembles have different properties and are individually driven
by near-resonant coherent laser fields, respectively. We in particular focus on the case in which the two subensembles interact via the dipole-dipole interaction~\cite{Agarwal,Ficek},
inducing cross-correlations. We find that the system does not assume a stationary state, but instead exhibits a stable population oscillation dynamics, even though
the atoms are damped by the coupling to the surrounding electromagnetic field reservoir. These long-term oscillations can be traced back to the dipole-dipole interaction,
and the oscillation frequency depends on the frequency difference of the laser fields driving the two ensembles. Further, the maximum excitation of the ensembles
sensitively depends on the inter-ensemble coupling and undergoes a collective ``jump'' if this coupling is slightly adjusted around a certain working point.
Interestingly, the jump is not between zero and full excitation of the ensemble but only involves few of the ensemble atoms.

Note that coupled ensembles of atoms have been studied intensely for various applications. For example, entanglement distribution over long distances
based on quantum repeaters using atomic ensembles as quantum memories \cite{rew} or Rydberg-blocked atomic ensembles \cite{ak} was successfully
demonstrated. Quantum networks \cite{qi} or quantum networks of atom clocks \cite{lukk} are further examples where coupled collections of atoms may have
an impressive impact. In this context, quantum teleportation between two remote atomic-ensemble quantum memory nodes, each composed of multiple rubidium
atoms ($\sim 100$) and connected by an optical fiber was already demonstrated in Ref.~\cite{pann}.

The paper is organized as follows. In Sec.~\ref{sec_eq} we describe our theoretical model, while in Sec.~\ref{sec_results} we analyze the obtained results via numerical simulation of the master equation. Section~\ref{sec_diss} discusses and summarizes our results.
\begin{figure}[t]
\centering
\includegraphics[width=\linewidth]{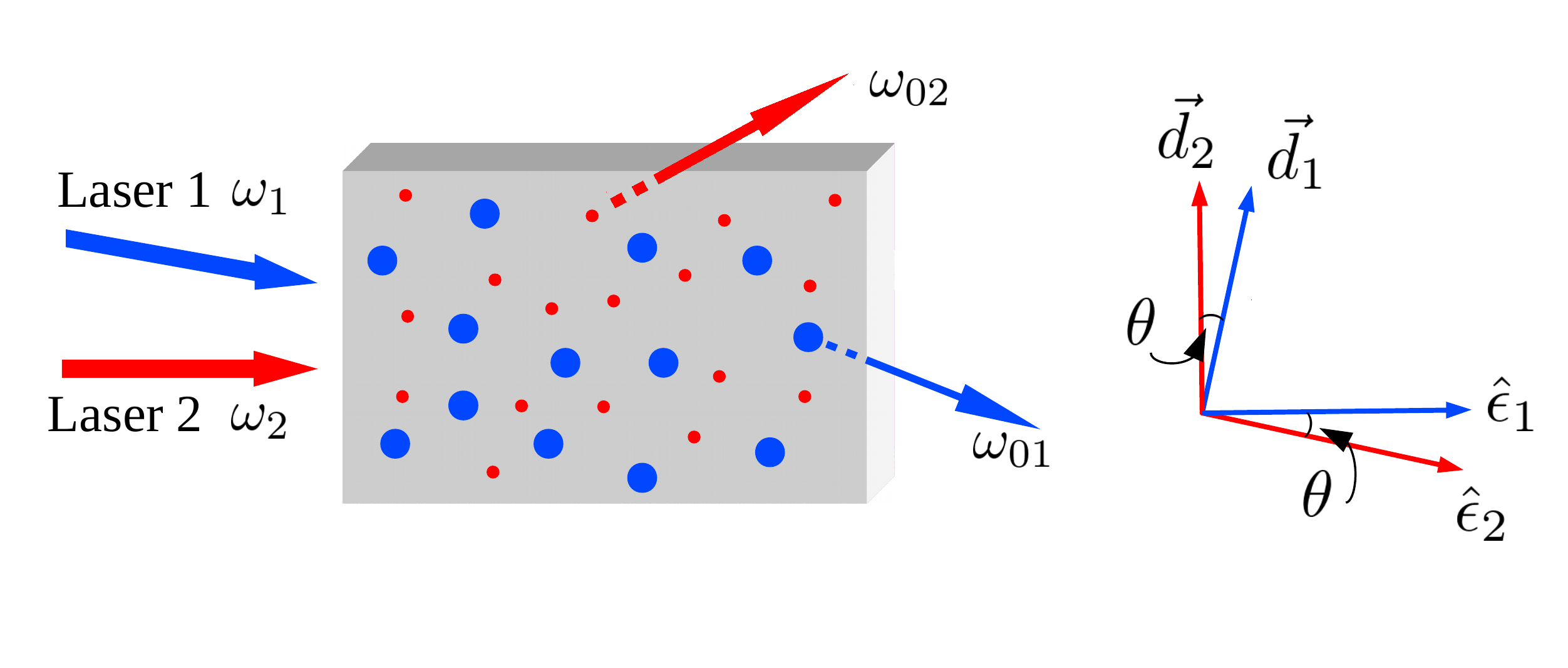}
\caption{ Schematic of the laser-driven atomic mixture. Big (blue) and small (red) dots represent two kinds of emitters (with transition
frequencies $\omega_{01}$ and $\omega_{02}$, respectively). Each of the atomic ensembles is pumped separately by laser 1 and 2 of
frequency $\omega_{1}$ and $\omega_{2}$, respectively, if $\vec{d}_1 \perp \hat \epsilon_2$ and $\vec{d}_2 \perp \hat \epsilon_1$. Similar inter-ensemble coupling can be realized when the atomic systems are spatially separated, e.g., via photon exchange mediated by  optical fibers or inter-cavities photon tunneling.}
\label{fig1}
\end{figure}

\section{Hamiltonian and Master Equation}\label{sec_eq}
We study an ensemble consisting of a mixture of two different types of many emitters (e.g., a sample with two kinds of atoms or nucleus). The emitters are pumped by two coherent  laser sources, as illustrated in Figure~\ref{fig1}. We use $\alpha, \alpha' \in \{1,2\}$ to label the two ensembles.
$\Omega_{\alpha \alpha'}=(\vec d_{\alpha'} \cdot \hat \epsilon _{\alpha} )E_{\alpha}/\hbar$ is the Rabi frequency of atoms in ensemble $\alpha'$ due to laser field $\alpha$, where $\vec d_{\alpha'}$ is the transition dipole moment, and $\hat \epsilon _{\alpha} E_{\alpha}$ is the vectorial
electric field amplitude which includes the polarization $\hat \epsilon _{\alpha}$.  By arranging
$\vec d_{\alpha'} \perp \hat \epsilon _{\alpha}$ (for $\alpha \neq \alpha'$) and $\vec d_{\alpha} \not\perp \hat \epsilon _{\alpha}$, as illustrated  in the right panel
of Fig.~\ref{fig1}, each subensemble is driven by a separate laser field. Thus in the following, we use $\Omega_{\alpha}$ to represent the Rabi frequency due
to laser $\alpha$ for the $\alpha$th sub-ensemble.

The Hamiltonian describing such a mixture in the interaction picture as well as rotating wave and dipole approximations is
(we set $\hbar=1$)
\begin{eqnarray}
H_{I}&=& \sum^{2}_{\alpha=1}\left [ (\Delta_{\alpha} - \beta_{\alpha}) S_{z\alpha}
+ \Omega_{\alpha}(\tilde S^{+}_{\alpha} + \tilde S^{-}_{\alpha})
+ \beta_{\alpha}\tilde S^{+}_{\alpha}\tilde  S^{-}_{\alpha} \right ]   \nonumber\\
&+& \beta_{12} \left (\tilde S^{+}_{1}\tilde S^{-}_{2} e^{i\phi(t)} +\tilde S^{+}_{2}\tilde S^{-}_{1} e^{-i\phi(t)}\right ).
\label{H}
\end{eqnarray}
Here $\tilde S^{\pm}_{\alpha}=S^{\pm}_{\alpha}e^{{\pm}i\varphi_{\alpha}}$, where $\varphi_{\alpha}$ is the phase of laser $\alpha$.
$S^{+}_{\alpha}=\sum^{N_{\alpha}}_{j=1}S^{+}_{j\alpha}$ [$S^{-}_{\alpha}=(S^{+}_{\alpha})^{\dagger}$] is the raising [lowering] collective operator
for the $\alpha$th atomic type and obeys the commutation relations of su(2) algebra, i.e,
\begin{align}
[S^{+}_{j\alpha},S^{-}_{j'\alpha'}] &=2S_{zj\alpha}\delta_{jj'}\delta_{\alpha\alpha'}\,,\\
[S_{zj\alpha},S^{\pm}_{j'\alpha'}]&=
\pm S^{\pm}_{j\alpha}\delta_{jj'}\delta_{\alpha\alpha'}\,,
\end{align}
where $N_{\alpha}$ is the particle number of type $\alpha$.
$\Delta_{\alpha}=\omega_{0\alpha}-\omega_{\alpha}$ is the corresponding detuning.
$\beta_{\alpha}$ gives the mean dipole-dipole interaction among emitters of $\alpha$th type.  $\beta_{12}=\eta \sqrt{\beta_{1}\beta_{2}}$ is the dipole-dipole
cross-coupling between these two kind of emitters with $0 \le \eta \le 1$ being a parameter characterizing it.
Finally,
\begin{align}
\phi(t)=-\phi_{0} + (\omega_{1} - \omega_{2})t,
\end{align}
with $\phi_{0} = \varphi_{1} - \varphi_{2}$ being
the phase difference between driving lasers.

The quantum dynamics of this laser-driven mixed multi-atom ensemble can be described by the following master equation
\begin{eqnarray}
\dot \rho = - i [H_{I}, \rho] - \mathbb{D},
\label{eq-final}
\end{eqnarray}
where, in the interaction picture, the Lindblad term is
\begin{eqnarray}
\mathbb{D}&=&\sum^{2}_{\alpha=1}\gamma_{\alpha}\left \{[\tilde S^{+}_{\alpha}, \tilde S^{-}_{\alpha}\rho] + [\rho \tilde S^{+}_{\alpha},\tilde S^{-}_{\alpha}]\right \}
\nonumber \\
&+& \eta \gamma_{12}\left\{[\tilde S^{+}_{1},\tilde S^{-}_{2}\rho]e^{i\phi (t)} + [\tilde S^{+}_{2},\tilde S^{-}_{1}\rho]e^{-i\phi (t)} + \textrm{ H.c.} \right\}. \nonumber \\
\label{dmp}
\end{eqnarray}
$\gamma_{\alpha}$ represents the spontaneous decay rate of atoms in ensemble $\alpha$, and $\gamma_{12}=\sqrt{\gamma_{1}\gamma_{2}}$.

In what follows, we apply the Holstein-Primakoff transformations \cite{HPT},
\begin{subequations}\label{HPT}
\begin{eqnarray}
\tilde S^{+}_{\alpha} &=& \sqrt{N_\alpha}a^{\dagger}_{\alpha}\sqrt{1 - \frac {a^{\dagger}_{\alpha}a_{\alpha}}{N_{\alpha}}},\\
\tilde S^{-}_{\alpha} &=& \sqrt{N_{\alpha}}\sqrt{1 - \frac {a^{\dagger}_{\alpha}a_{\alpha}}{N_{\alpha}}}a_{\alpha},  \\
S_{z\alpha}&=& a^{\dagger}_{\alpha}a_{\alpha} - N_{\alpha}/2,
\end{eqnarray}
\end{subequations}
where $a^{\dagger}_{\alpha}$ and $a_{\alpha}$ are the creation and the annihilation operators associated to $\alpha$-type emitters and obey
the standard bosonic commutation relations. Assuming low excitation numbers, i.e., $\langle a^{\dagger}_{\alpha}a_{\alpha}\rangle /N_{\alpha} \ll 1$, one can
expand the collective operators $\tilde S^{\pm}_{\alpha}$ in Eq.~(\ref{HPT}) up to the first-order terms of $a^{\dagger}_{\alpha}a_{\alpha}$ in this small
parameter,
\begin{subequations}
\label{spm}
\begin{eqnarray}
\tilde S^{+}_{\alpha} \approx \sqrt{N_{\alpha}}a^{\dagger}_{\alpha}\left(1 - \frac {a^{\dagger}_{\alpha}a_{\alpha}}{2N_{\alpha}}\right),\\
\tilde S^{-}_{\alpha} \approx \sqrt{N_{\alpha}}\left (1 - \frac {a^{\dagger}_{\alpha}a_{\alpha}}{2N_{\alpha}}\right )a_{\alpha}.
\end{eqnarray}
\end{subequations}
For the term $\tilde  S^{+}_{\alpha}\tilde S^{-}_{\alpha}$
in the master equation (\ref{eq-final}), we can calculate directly from Eq.~(\ref{HPT}) that
\begin{eqnarray}
\tilde S^{+}_{\alpha}\tilde S^{-}_{\alpha} = N_{\alpha}a^{\dagger}_{\alpha}a_{\alpha} - a^{\dagger 2}_{\alpha}a^{2}_{\alpha}.
\label{sdd}
\end{eqnarray}

The mean numbers of excitations in each sub-ensembles are calculated from the density matrix (\ref{eq-final}) as
\begin{eqnarray}
\langle N_{\mathrm e\alpha} \rangle= Tr\{ a^{\dagger}_{\alpha}a_{\alpha} \rho\}= \sum_{m_{1}=0}\sum_{m_{2}=0}m_{\alpha}P_{m_{1},m_{2};m_{1},m_{2}},
\label{ne}
\end{eqnarray}
where   the elements of the density matrix are given by
\begin{eqnarray*}
P_{m_{1},m_{2};n_{1},n_{2}} =  \langle m_{1},m_{2}|\rho|n_{1},n_{2} \rangle.
\end{eqnarray*}
Employing the relations (\ref{spm},\ref{sdd}) in the master equation (\ref{eq-final}) enables one to numerically solve the problem. In the next Section, we discuss the results.

\section{Numerical results}\label{sec_results}
\begin{figure} [h]
\centering
\includegraphics[width=\linewidth]{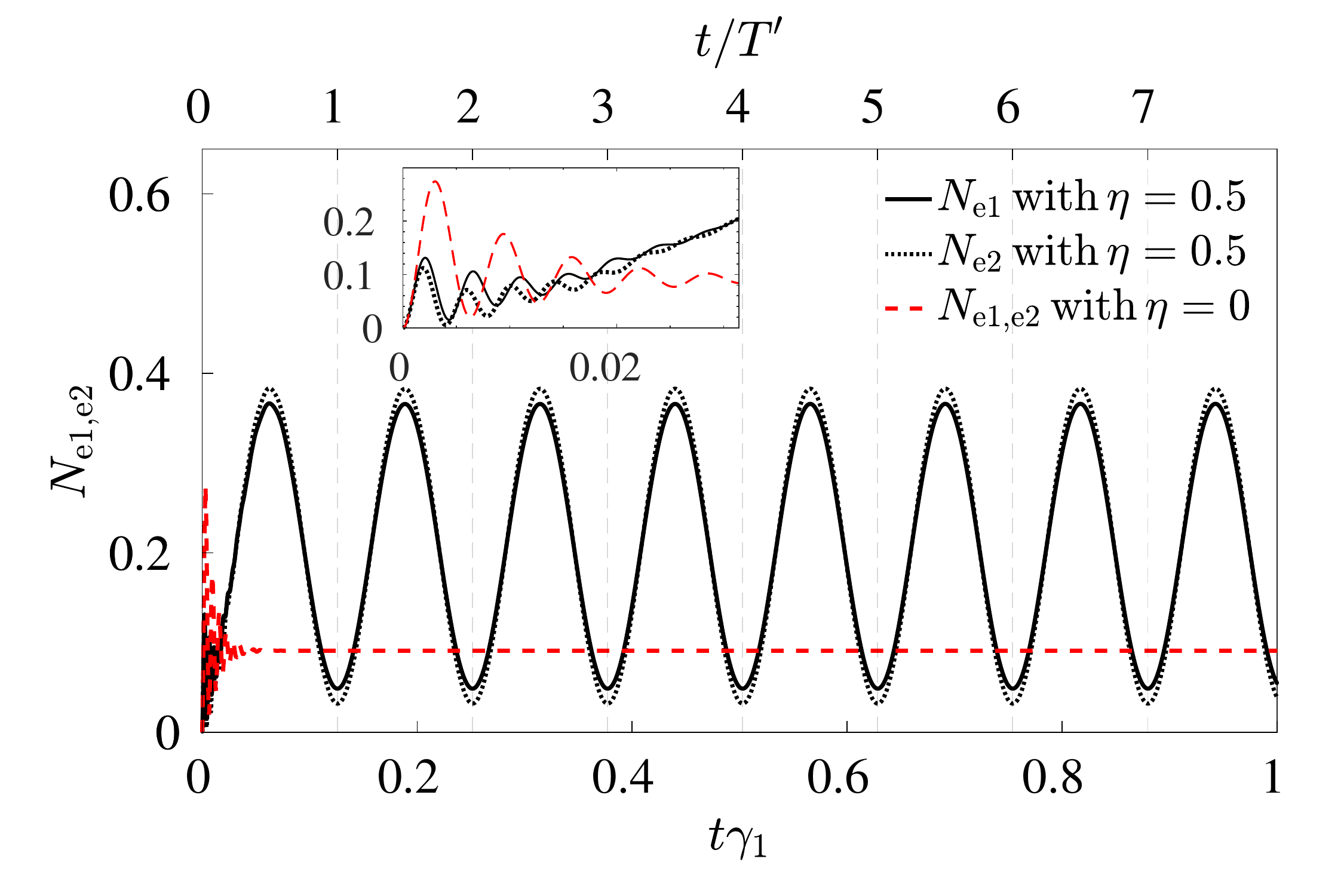}
\caption{ Evolution of the excited-state populations $N_{\mathrm e1}$ and $N_{\mathrm e2}$ of the two  laser-driven ensembles with and without cross-coupling terms.
The parameters are in units of $\gamma_{1}$, which are, $\gamma_{2}=1$, $\Delta_{1,2}=0$, $\beta_{1,2}=10$,
$\Delta\omega=|\omega_{1} - \omega_{2}| = 50$, $\Omega_{1,2}=30$, and $\phi_{0}=0$. The numbers of atoms in the two ensembles are $N_{1,2}=100$. The upper \textit{t} - axis is in units of $T'=2\pi/\Delta\omega=2\pi/50$, in order to illustrate the dependence on the frequency difference of the driving laser fields. The black-solid (dotted) curve represents $N_{\mathrm e1}$  ($N_{\mathrm e2}$) with the cross-coupling rate
$\eta=0.5$, whereas the red-dashed curve is for $\eta=0$. The inset shows the population evolution in a short time.}
\label{fig2}
\end{figure}

We start by evaluating the mean number of excitations in the atomic mixture given by Eq.~(\ref{ne})
in the presence $(\eta \not = 0)$ or absence $(\eta = 0)$ of cross-couplings among different atomic species. The red-dashed curve in Fig.~\ref{fig2} shows the
results without cross-couplings. We find that, as expected, the two subensembles evolve independently, and the mean numbers of excitation $N_{\mathrm e1}$ and $N_{\mathrm e2}$ have analogous time dependence because of the symmetrical parameters chosen in our numerical calculations.
After only few oscillations, the system reaches its steady state, on a time scale of the inverse decay rates. The frequencies of these oscillations do not correspond to the collective Rabi frequency $\sqrt{N_\alpha}\Omega_\alpha$ \citep{CRF,CRF2} because our system is not described by a collective state with only one excitation since we have taken the possibility of multiple excitations into consideration.
In contrast, if cross-couplings between the two subensembles exist, no steady steady state is assumed. Instead, the oscillations in the excitation numbers as function of time persist. As an example, the black curves in Fig.~\ref{fig2} show the results when $\eta=0.5$. In the following, we study these long-term oscillations in more detail.

\begin{figure} [h]
\centering
\includegraphics[width=\linewidth]{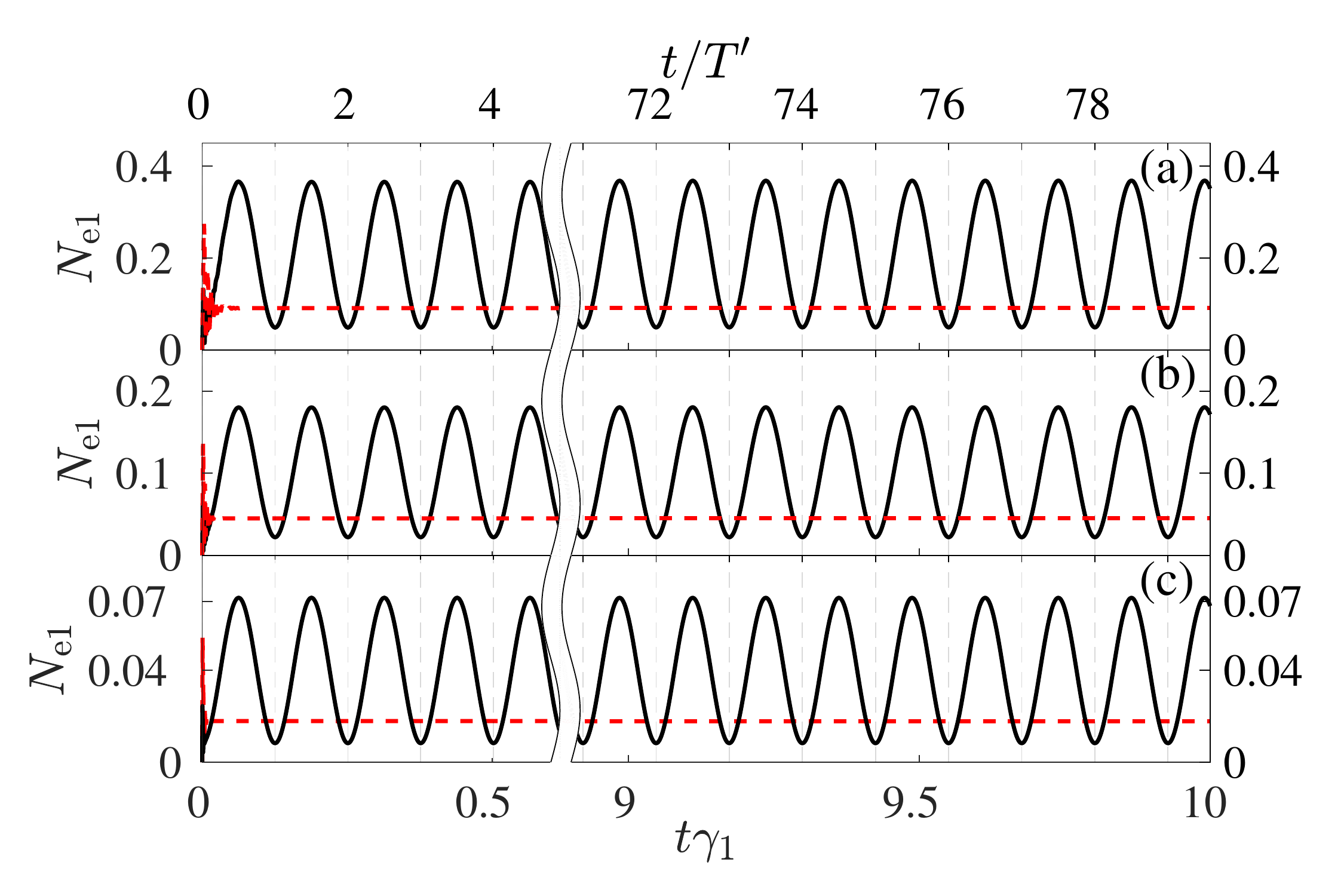}
\caption{ The long-time evolution of the excitation number $N_{\mathrm e1}$ with different particle numbers: (a) $N_{1,2}=100$, (b) $N_{1,2}=200$, and (c) $N_{1,2}=500$.
The black-solid curves represent $N_{\mathrm e1}$ with the cross-coupling rate
$\eta=0.5$, whereas the red-dashed ones are for $\eta=0$. Other parameters are the
same as in Fig.~\ref{fig2}.}
\label{fig3}
\end{figure}

From the Hamiltonian in Eq.(\ref{H}) and the master equation in Eq.(\ref{eq-final}) we find that the cross-couplings between the two ensembles induce a time-dependent oscillating term $e^{\pm i\phi (t)} $, which is
proportional to $\eta$. It is this term that is responsible for the long-time oscillations in the  populations.
The frequency of these oscillations is $|\omega_{1}-\omega_{2}|=\Delta\omega$, as can be seen from the scaling with $T'=2\pi/\Delta\omega$ in the upper \textit{t}-axis in Fig.~\ref{fig2}. For comparison, the lower \textit{t}-axis is in units of inverse $\gamma_{1}$. The inset of Fig. 2 shows that in the presence of cross-coupling
the two ensembles synchronize within a short time duration.
The dependence of the long-time oscillation frequency on the difference of the laser
frequencies is further confirmed by detuning one of the laser fields.
Due to the ensemble nature of the two coupled subsystems, the long-term dynamics induced by the cross-couplings depends on the particle numbers. To illustrate this, we show the long-time excitation dynamics for different particle numbers $N_{1,2}=100,~200,~500$ in Figs.~\ref{fig3}(a)-\ref{fig3}(c), respectively. While the oscillation frequency is not affected by the particle number, the amplitude of the oscillations is.

Due to the expansion in $a^{\dagger}_{\alpha}a_{\alpha}$, our results are valid as long as $\langle a^{\dagger}_{\alpha}a_{\alpha}\rangle/N_{\alpha} \ll 1$, where $\alpha \in \{1,2\}$. In the following, we consider larger atomic ensembles. If further the driving laser fields are weak enough, one may also ignore the first-order terms in $a^{\dagger}_{\alpha}a_{\alpha}$ when expanding the square roots in Eq.~(\ref{HPT}), such that $\tilde S^{+}_{\alpha} \approx \sqrt{N_{\alpha}}a^{\dagger}_{\alpha}$ and $\tilde S^{-}_{\alpha}\approx \sqrt{N_{\alpha}}a_{\alpha}$. This stronger approximation considerably accelerates the numerical calculation and has been numerically verified by comparing with the numerical results using Eq.~(\ref{spm}) in a shorter time range up to $T'$.

\begin{figure} [h]
\centering
\includegraphics[width=\linewidth]{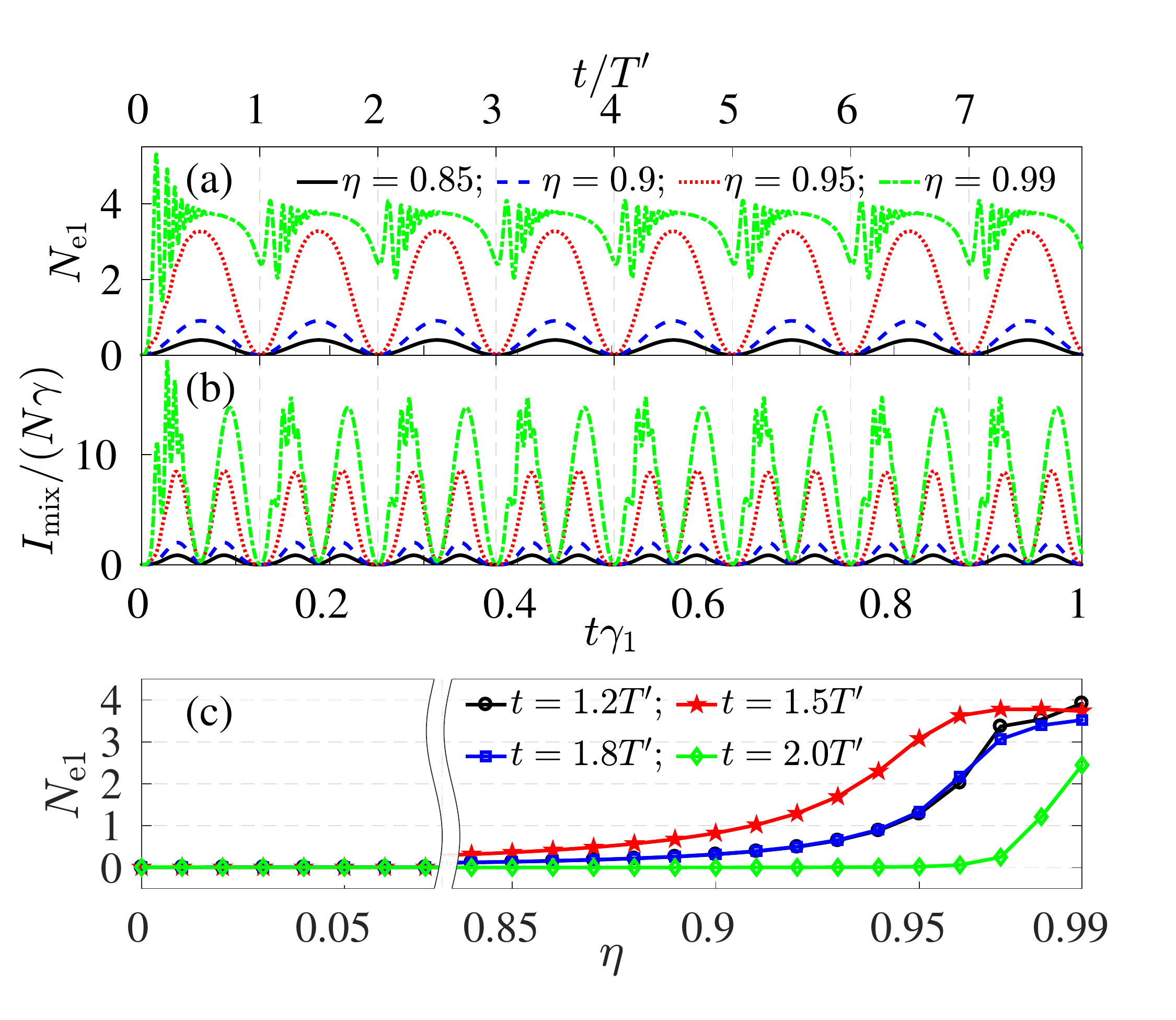}
\caption{ Dependence of the excitation number $N_{\mathrm e1}$ (a) and the intensity of the spontaneously
scattered photons $I_{\mathrm{mix}}/(N\gamma)$ (b) on the scaled time. (c) $N_{\mathrm e1}$ at certain moments in time as a function of the cross-coupling rate $\eta$. The number of atoms is $N_{1,2}=1000$. Other parameters are the same as in Fig.~\ref{fig2}.}
\label{fig4}
\end{figure}

\begin{figure} [h]
\centering
\includegraphics[width=\linewidth]{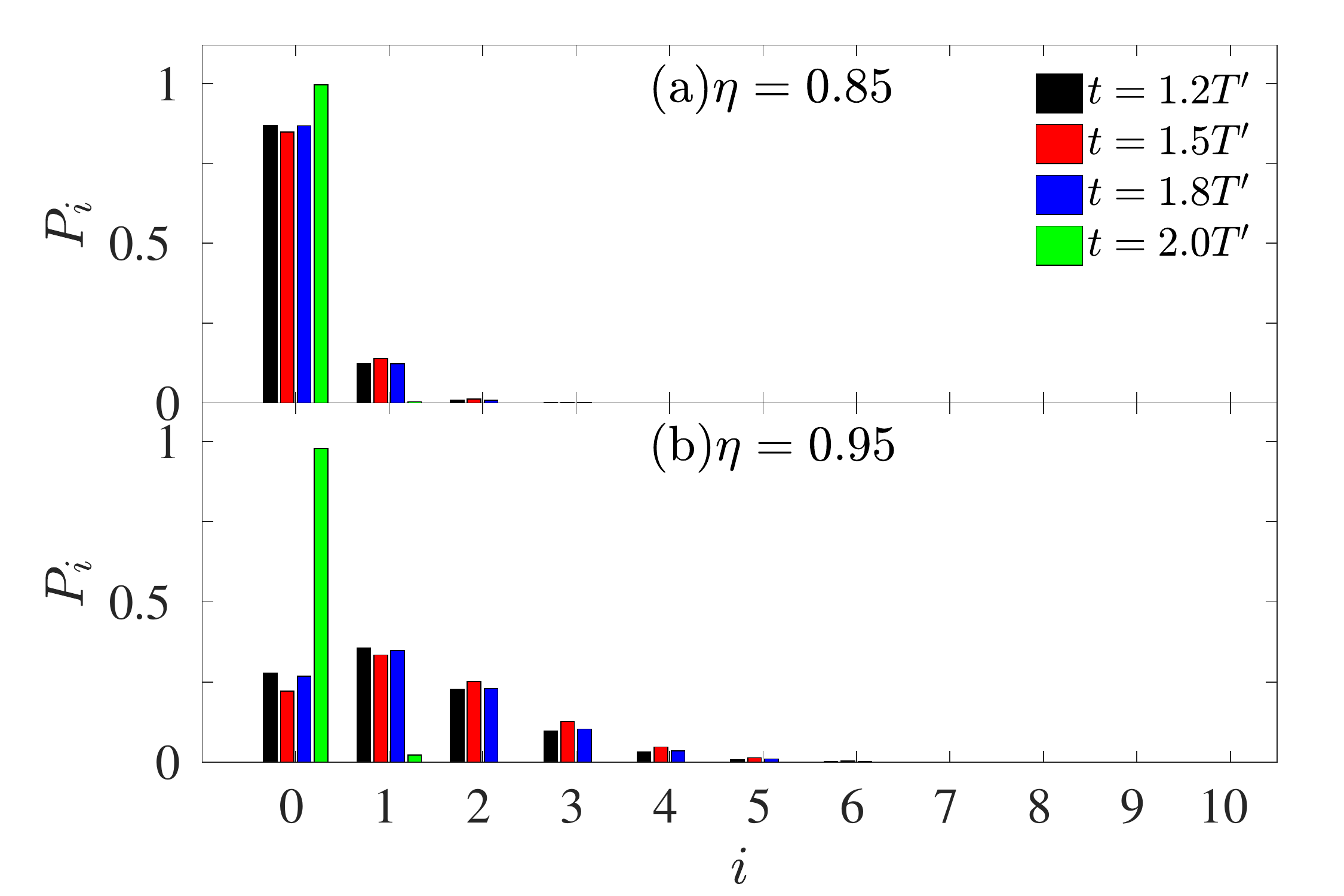}
\caption{ Probability to excite $i$ atoms inside the first ensemble with (a) $\eta=0.85$ and (b) $\eta=0.95$. The number of atoms is
$N_{1,2}=1000$ while all other parameters are the same as in Fig.~\ref{fig2}.}
\label{fig5}
\end{figure}

Results  with $N_1=N_2=1000$ are shown in Fig.~\ref{fig4} and Fig.~\ref{fig5}.
Figure~\ref{fig4} illustrates the dependence on the cross-coupling rate $\eta$. In Fig.~\ref{fig4}(a), we find that the degree of excitation grows with increasing $\eta$ for fixed laser intensities. However, while initially also the oscillation amplitude in the excitation number increases with $\eta$, the dynamics becomes more complex and saturates for $\eta$ approaching unity.
With strong cross-coupling, rapid suboscillations of the populations
become visible, e.g., in the green (dash-dotted) curve when $\eta=0.99$.
The origin of these oscillations is a breakdown of adiabaticity. The
system tries to evolve into a steady state, where no oscillations occur.
However, due to the cross-couplings, the temporal variation of $\phi(t)=-\phi_0+(\omega_1-\omega_2)t$
moves the system out of the steady state. The rapid oscillations then
occur if the evolution of the system cannot adiabatically follow the
changes induced by $\phi(t)$. This is more likely at large $\eta$ since
then some of the system states become slowly-decaying trapping states,
such that the evolution into the steady state becomes slower.
Note that a quantitative analysis of adiabaticity would require the determination of the properties of the relevant sub-radiant collective states, which is challenging due to the complexity of our system.
Similarly,
oscillations also occur for lower $\eta$ if the rate of change of
$\phi(t)$ is increased. Note that the period of the rapid evolution is
governed by a collective Rabi frequency of the ensemble.
The role of trapping states can further be elucidated via the total
intensity of the spontaneously
scattered photons as function of time, given by
\begin{align}
\frac{I_{\mathrm{mix}}}{N\gamma}&=\langle
a_1^{\dagger}a_1\rangle+\langle a_2^{\dagger}a_2\rangle +\eta\langle
a_1^{\dagger}a_2\rangle e^{-i\phi(t)}+\eta\langle
a_1a_2^{\dagger}\rangle e^{i\phi(t)}\\  \nonumber
&=N_{e1}+N_{e2}\\  \nonumber
&+\eta
e^{-i\phi(t)}\sum_{m_1}\sum_{m_2}\sqrt{m_1}\sqrt{m_2+1}P_{m_1
-1,m_2+1;m_1,m_2}\\ \nonumber
&+\eta
e^{i\phi(t)}\sum_{m_1}\sum_{m_2}\sqrt{m_1+1}\sqrt{m_2}P_{m_1+1,m_2-1;m_1,m_2},
\end{align}
where $N_1=N_2=N$, and $\gamma_1=\gamma_2=\gamma$. Results are shown in
Fig.~\ref{fig4}(b). It can be seen that the total scattered intensity
has a minimum where the excited state populations are maximal,
demonstrating that the system is in a trapped state.
To elucidate the role of $\eta$ further, we investigate the degree of excitation at certain moments in  time  throughout the dynamics, see Fig.~\ref{fig4}(c). At low values of $\eta$, the system is almost in the ground state, which is predictable from the results in Figs.~\ref{fig3}. Hoowever, starting from a threshold value, the degree of excitation quickly increases with growing $\eta$ and then saturates. The threshold and the steepness of the ``jump'' in excitation depends on the time of detection.
Note that collective population jumps as functions of a control parameter are well known in pumped
multi-atom ensembles \cite{andreev,puri,jump}. However, here, not the entire ensemble but only a small fraction of the ensemble sensitively depends on an external parameter.

We have also the excitation statistics as function of $\eta$ across the jump. In Fig.~\ref{fig5},  the probability to excite $i$ atoms in the first atomic ensemble is shown, i.e., $P_{i}=\sum_{n_{2}}\langle i,n_{2}|\rho|i,n_{2}\rangle$. In Fig.~\ref{fig5}(a), it can be seen that below the threshold, the system is mostly in the ground state. Figure \ref{fig5}(b) shows that above the threshold, multiple excitations  occur. The atomic distribution $P_i$ broadens with increasing $\eta$, but remains Poissonian.

\section{Discussion and summary}\label{sec_diss}

We found that the cross-couplings between the two subensembles may prevent the system from entering a stationary state. The cross-coupling rate $\eta$ further acts as a control parameter for the ensemble excitation. At small $\eta$, the ensemble population is low in the studied parameter range, while beyond a certain threshold, the excitation quickly raises with increasing $\eta$ and then saturates to a small fraction of the total number of atoms.

The long-time oscillations can be understood by noting that the cross-coupling acts as an additional pumping source for each of the two subensembles, with a frequency that differs in magnitude by $\Delta \omega= |\omega_{1}-\omega_{2}|$ from the applied laser field. This bichromatic pumping leads to the nonstationary long-time dynamics. Thus the frequency of the long-time oscillations is determined by the frequency of the additional pump, i.e. $\Delta \omega$, and the amplitude is affected by the strength of the additional pump, namely $\eta$. This interpretation is similar to that for the long-time oscillations found in a two-atom few-level system~ \cite{ekmk}. However, in the latter case, the long-time dynamics was determined in a geometry-dependent configuration due to the multi-level nature of the two atoms. In contrast, here, cooperative effects induce a dependence on the number of atoms in the ensemble and to the ``jump'' in the atom excitation as function of $\eta$.

It is challenging to find the analytical solution of the master equation for our mixed multi-atom system. However, we can analyze under which conditions the long-time oscillations occur. For this, we assume a large system with low excitation, such that one can keep only  the zeroth-order terms in $a^{\dagger}_{\alpha}a_{\alpha}$ in the expansion of the Holstein-Primakoff transformations. In this case, the  interaction Hamiltonian takes the form
\begin{eqnarray}
H_{I} &=& \sum^2_{\alpha=1}\left [ \bar \Delta_{\alpha}a^{\dagger}_{\alpha}a_{\alpha} +  \bar \Omega_{\alpha}(a^{\dagger}_{\alpha} + a_{\alpha}) \right ] \nonumber \\
&+& \beta_{12}\sqrt{N_{1}N_{2}} \left (a^{\dagger}_{1}a_{2} e^{i\phi (t)} + a^{\dagger}_{2}a_{1}e^{-i\phi (t)} \right ) \,,
\label{simH}
\end{eqnarray}
where $\bar \Delta_{\alpha}=\Delta_{\alpha} + (N_{\alpha} - 1)\beta_{\alpha}$ is the effective detuning, whereas
$\bar \Omega_{\alpha}= \Omega_{\alpha}\sqrt{N_{\alpha}}$ is the collective Rabi frequency and $\beta_{12}=\eta \sqrt{\beta_{1}\beta_{2}}$.
The long-time oscillations arise from the term proportional to $\beta_{12}$. This term can be neglected if it is rapidly oscillating, i.e., if  $|\omega_{1}-\omega_{2}|$ is significant larger than $\beta_{12}\sqrt{N_{1}N_{2}}$, such that the term averages out from the dynamics. From Eq.~(\ref{dmp}), we further find that for the spontaneous emission part, the cross-correlations can be ignored if $|\omega_{1}-\omega_{2}| \gg \eta \sqrt{\gamma_{1}\gamma_{2}N_{1}N_{2}}$.
For our parameters, these two conditions are not fulfilled, such that the oscillations are observed.

The initial phase difference $\phi_0$ is introduced into the equations of motion via the cross-coupling terms. However, we found that $\phi_0$ does not affect the amplitude or frequency of the long-time oscillations, or the excitation jumps.

Our results suggest diverse applications. For example, we found that the population sensitively depends on the system parameters in a strongly nonlinear way. This provides a handle to characterize the coupling between the two ensembles, in particular, close to the desirable maximum coupling $\eta\approx 1$. As shown in Fig.~\ref{fig4}, the magnitude of the coupling $\eta$ crucially determines the population of the ensembles, and around a certain working point, small changes in $\eta$  manifest themselves in substantial modifications of the atom excitation.

Another possible application is the synchronization of the dynamics of two initially independent and spatially separated many-atom ensembles with different transition frequencies~\cite{l3}. As shown in the inset of Fig.~\ref{fig2}, initially, the two ensembles have different time evolutions of the excited-state population. However, after a short time, the systems evolve into the long-time oscillatory dynamics, in which the excitations of the two ensembles as function of time are synchronized. This way, it would, for example, be possible to reliably excite both ensembles to the same fraction of the maximum possible excitation. It is important to note that this synchronization is robust in that it does not depend on the symmetric parameter choice in Fig.~\ref{fig2}, but also occurs e.g., with different Rabi frequencies and laser detunings for the two ensembles. Such a synchronization could be detected, e.g., via fluorescence (see, also, \cite{l1} and \cite{l2}). It could also serve as a method to monitor the dynamics
of one ensemble with a second ancilla ensemble.

\textbf{Acknowledgment.}
L.J. and M.M.A. acknowledge the financial support given by the German Federal Ministry of Education and
Research, grant No. 01DK13015, and the Academy of Sciences of Moldova, grants No. 13.820.05.07/GF and 15.817.02.09F.
Furthermore we are grateful for the hospitality of the Theory Division of the Max Planck Institute for Nuclear Physics from Heidelberg, Germany.



%

\end{document}